\documentclass[aps,prb,a4paper,10pt,twocolumn,showpacs,floatfix,superscriptaddress,preprintnumbers,longbibliography]{revtex4-2}
\makeatletter

\newcommand{\Rmnum}[1]{\expandafter\@slowromancap\romannumeral #1@}
\makeatother
\usepackage{float}
\usepackage{indentfirst}
\usepackage{dcolumn,graphicx,color,booktabs,microtype,afterpage}
\usepackage{amssymb}
\usepackage{amsmath}
\usepackage[charter,greekuppercase=italicized]{mathdesign}
\usepackage{sidecap}
\usepackage[mathlines]{lineno}
\usepackage{bm}
\graphicspath{{./}{figure/}}
\renewcommand{\tablename}{Table}
\makeatletter\renewcommand{\fnum@figure}[1]{\figurename~\thefigure.~}\makeatother
\makeatletter\renewcommand{\fnum@table}[1]{\tablename~\thetable.}\makeatother

\newcount\hh \newcount\mm
\hh=\time \divide\hh by 60
\mm=\hh \multiply\mm by 60 \mm=-\mm
\advance\mm by \time
\def\now{\number\hh:\ifnum\mm<10{}0\fi\number\mm}

\usepackage[colorlinks,plainpages=false,linkcolor=blue,urlcolor=blue,citecolor=blue,pdfpagemode=UseNone,pdfstartview=FitBH]{hyperref}

\newcommand{\tcr}[1]{\textcolor{black}{#1}}
\newcommand{\tcb}[1]{\textcolor{blue}{#1}}

\begin{document}

\makeatletter\renewcommand{\ps@plain}{%
\def\@evenhead{\hfill\itshape\rightmark}%
\def\@oddhead{\itshape\leftmark\hfill}%
\renewcommand{\@evenfoot}{\hfill\small{--~\thepage~--}\hfill}%
\renewcommand{\@oddfoot}{\hfill\small{--~\thepage~--}\hfill}%
}\makeatother\pagestyle{plain}

\preprint{\textit{Preprint: \today, \now}} %For internal use only, do not distribute.}}
%\linenumbers
%\title{Ideal two-dimensional frustrated magnetism in hexagonal perovskite Ba$_2$La$_2$MnTe$_2$O$_{12}$}
%\title {\tcr{Two-dimensional frustrated magentism in a three-dimensional hexagonal perovskite}}

\title{Strongly frustrated 2D magnetism in a 3D hexagonal perovskite}

	%Ideal Two-Dimensional Magnetism in Hexagonal Perovskites Ba$_2$La$_2$MnTe$_2$O$_{12}$}
	%Unique Magnetic Structure in the Ideal Two-Dimensional Triangular Lattice Antiferromagnet Ba$_2$La$_2$MnTe$_2$O$_{12}$}
%
\author{Bocheng\ Yu}\thanks{These authors contributed equally.}
\affiliation{School of Physics, East China Normal University, Shanghai 200241, China}
\author{Otkur\ Omar}\thanks{These authors contributed equally.}
\affiliation{Department of Physics, University of Science and Technology of China, Hefei, 230026 Anhui, China}
\author{Songtai\ Lv}\thanks{These authors contributed equally.}
\affiliation{School of Physics, East China Normal University, Shanghai 200241, China}
\author{Long\ Ma}
\affiliation{Anhui Province Key Laboratory of Condensed Matter Physics at Extreme Conditions, High Magnetic Field Laboratory, Chinese Academy of Sciences, Hefei 230031, China}
\author{Zhengcai\ Xia}
%\affiliation{Huazhong University of Science and Technology, Wuhan 430074, China}
\affiliation{Wuhan National High Magnetic Field Center and School of Physics, Huazhong University of Science and Technology, Wuhan 430070, China}

\author{Jing\ Meng}
\affiliation{School of Physics, East China Normal University, Shanghai 200241, China}
\author{Yanran\ Yang}
\affiliation{School of Physics, East China Normal University, Shanghai 200241, China}
\author{Jie\ Ma}
\affiliation{School of Physics and Astronomy, Shanghai Jiao Tong University, Shanghai 200240, China}
\author{Yang\ Xu}
\affiliation{School of Physics, East China Normal University, Shanghai 200241, China}
\affiliation{Key Laboratory of Polar Materials and Devices (MOE), East China Normal University, Shanghai 200241, China}
\author{Qingfeng\ Zhan}
\affiliation{School of Physics, East China Normal University, Shanghai 200241, China}
\affiliation{Key Laboratory of Polar Materials and Devices (MOE), East China Normal University, Shanghai 200241, China}
\author{Vladimir\ Yu.\ Pomjakushin}
\affiliation{PSI Center for Neutron and Muon Sciences CNM, CH-5232 Villigen PSI, Switzerland}
\author{Haiyuan\ Zou}
\affiliation{School of Physics, East China Normal University, Shanghai 200241, China}
\affiliation{Key Laboratory of Polar Materials and Devices (MOE), East China Normal University, Shanghai 200241, China}
\author{Shang\ Gao}\email[Corresponding author:\\]{sgao@ustc.edu.cn}
\affiliation{Department of Physics, University of Science and Technology of China, Hefei, 230026 Anhui, China}
\affiliation{Key Laboratory of Strongly-Coupled Quantum Matter Physics, Chinese Academy of Sciences, School of Physical Sciences,
	University of Science and Technology of China, Hefei, Anhui 230026, China}
\author{Toni\ Shiroka} \email[Corresponding authors:\\]{tshiroka@phys.ethz.ch}
\affiliation{PSI Center for Neutron and Muon Sciences CNM, CH-5232 Villigen PSI, Switzerland}
\affiliation{Laboratorium f\"ur Festk\"orperphysik, ETH Z\"urich, CH-8093 Z\"urich, Switzerland}
\author{Tian\ Shang}\email[Corresponding authors:\\]{tshang@phy.ecnu.edu.cn}
\affiliation{School of Physics, East China Normal University, Shanghai 200241, China}
\affiliation{Key Laboratory of Polar Materials and Devices (MOE), East China Normal University, Shanghai 200241, China}
%\affiliation{Chongqing Key Laboratory of Precision Optics, Chongqing Institute of East China Normal University, Chongqing 401120, China}

%
\begin{abstract}
Exotic quantum phenomena are often found to occur in spin systems that exhibit low-dimensional magnetism. By combining nuclear magnetic resonance, neutron scattering, and muon-spin spectroscopy ({\textmu}SR) techniques, we report a rare instance of strongly frustrated two-dimensional (2D) magnetism in a three-dimensional (3D) hexagonal perovskite.
Here, Ba$_2$La$_2$MnTe$_2$O$_{12}$, a triangular-lattice magnet, is shown to undergo a magnetic transition at $T_\mathrm{N} \approx 4.4$\,K, below which the manganese moments form a 120$^{\circ}$ AFM order within the $ab$-plane, while staying disordered along the $c$-axis. This exotic ground state, which exhibits ideal 2D magnetism, is highly consistent with the persistently strong spin fluctuations and the large internal field distributions revealed by zero-field {\textmu}SR.
Further, the 2D magnetism also leads to a significant frustration,
much larger than that of most known magnetically-ordered frustrated systems. Our work on Ba$_2$La$_2$MnTe$_2$O$_{12}$ not only challenges the interpretations of magnetic order in other 3D hexagonal perovskites, but it also provides insight into how the dimensionality affects the exotic magnetic states.
\end{abstract}

%\keywords{Unconventional supeconductivity, muon spin rotation/relaxation, nuclear magnetic resonance}

\maketitle\enlargethispage{3pt}

\vspace{-5pt}
\section{\label{sec:Introduction}Introduction}\enlargethispage{8pt}

Low-dimensional spin systems with magnetic frustration are ideal platforms for realizing exotic quantum states, such as quantum spin liquids~\cite{balents2010,broholm2020,zhou2017,savary2017}, spin ice~\cite{skjaervo2020,gingras2014}, spin supersolids~\cite{heidarian2005,yamamoto2014,sellmann2015}, Bose-Einstein condensation~\cite{matsubara1956,giamarchi2008,zapf2014}, etc.
Most of these low-dimensional %spin
systems host magnetic lattices with geometric frustration (also known as magnetic frustration), including kagome and triangular lattices~\cite{paddison2017,li2020,cheng2011,zhou2012,gao2022,sheng2025,bordelon2019,xiang2024,ding2018,fak2012,helton2007,singh2012,matsumoto2024}. %zhong2020,lefrancois2016,sears2015}.
The latter systems are among  the best models for studying the occurrence of exotic quantum states~\cite{paddison2017,li2020,cheng2011,zhou2012,gao2022,sheng2025,bordelon2019,xiang2024}. 

\begin{figure*}[!htb]
	\centering
	\includegraphics[width=0.99\textwidth,angle=0]{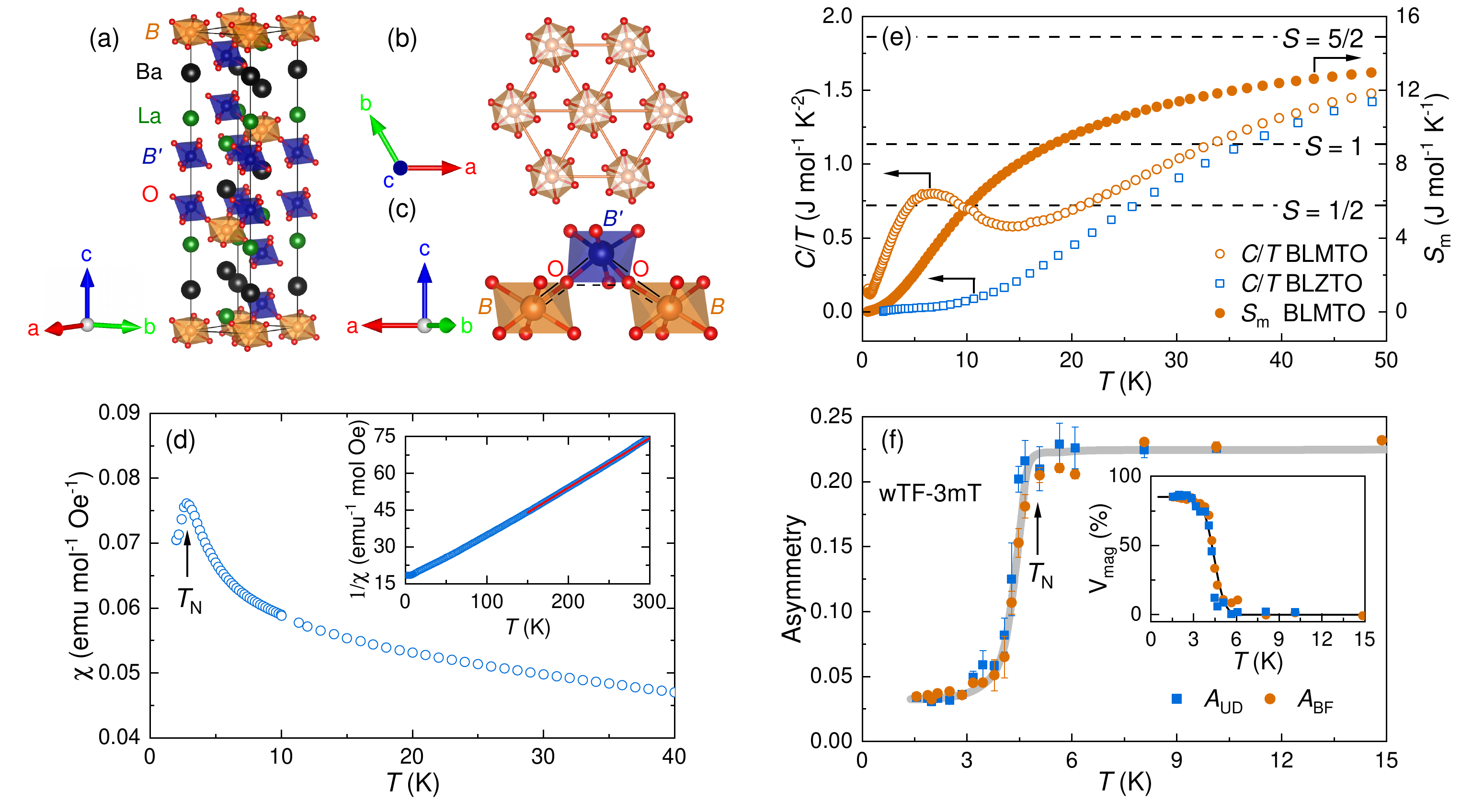}
	\vspace{-2ex}%
	\caption{\label{fig:Cp} \tcr{(a) Crystal structure of Ba$_2$La$_2$$BB’_2$O$_{12}$ ($B$ = Co, Ni, Mn; $B’$ = W, Te, Re). (b) Triangular sublattice of the magnetic $B$ ions. (c) Pathways of the $B$-O-$B’$-O-$B$ (black solid lines) and $B$-O-O-$B$ (black dashed lines) superexchange interactions. (d) Temperature-dependent magnetic susceptibility $\chi(T)$ for BLMTO collected in a field of 50\,mT. The inset shows the inverse susceptibility 1$/\chi(T)$. The red solid line represents a fit to the Curie-Weiss law. (e) Temperature-dependent specific heat $C(T)/T$ (left-axis) and magnetic entropy $S_\mathrm{m}(T)$ (right-axis) of BLMTO. The magnetic specific heat was obtained by subtracting the specific heat
	of the nonmagnetic Ba$_2$La$_2$ZnTe$_2$O$_{12}$ (BLZTO) (empty circles and squares, resp.,  see details in Fig.~S4~\cite{Supple}) from the raw data. Dashed lines mark the entropy values $R\ln(2S + 1)$, with $S$ = 1/2, 1, and 5/2, respectively. 
	(f) Temperature dependence of nonmagnetic asymmetry $A_\mathrm{NM}$ of wTF-{\textmu}SR spectra for BLMTO, obtained in a 3-mT transverse field.  Data from the up-down (UD) and backward-forward (BF) detectors are highly consistent. The inset shows the magnetic volume fraction vs temperature. The solid line is a fit to the phenomenological relation 
	$V_\mathrm{mag}(T) = V_\mathrm{mag}(0)\;\frac{1}{2} [1 - \mathrm{erf}(\frac{T-T_\mathrm{N}}{\sqrt{2}\Delta T})]$, where $\Delta T$, $V_\mathrm{mag}(0)$ and erf($T$) are the transition width, the zero-temperature magnetic volume, and the error function, respectively. }
	}
\end{figure*}

To the triangular lattice magnets belong also the hexagonal perovskites $A_4BB'_2$O$_{12}$ [see Figs.~\ref{fig:Cp}(a) and (b)], with $A$ = Sr, Ba, La; $B$ = Co, Ni, Mn; and $B'$ = Re, W, Te~\cite{kojima2018,yu2023,saito2019,rawl2017,doi2017,park2024}.
Depending on the stoichiometry of the $B'$ site, the $A$ site may be occupied entirely by Ba$^{2+}$, Sr$^{2+}$, or La$^{3+}$ ions, or by a combination of them~\cite{rawl2017}. In this structure, the $B$O$_6$ octahedra share their corners with the $B'$O$_6$ octahedra via oxygen atoms, which leads to two types of superexchange interaction paths, i.e., ferromagnetic (FM) $B$--O--$B'$--O--$B$ and antiferromagnetic (AFM)  $B$--O--O--$B$ [see Fig.\ref{fig:Cp}(c)]~\cite{rawl2017}. 
%	\tcy{[It is necessary to remind that although there is indeed such a conclusion in the cited literature (rawl2017), when $B'$ = Te, the interactions of both paths are AFM.]} 
The competition between these FM and AFM interactions eventually determines the magnetic ground state of $A_4BB'_2$O$_{12}$, which can be tuned by physical pressure or by chemical substitutions (see Table~S1 in the Supplementary Materials~\cite{Supple} and also Refs.~\onlinecite{Li2022,Nishino1996,Corboz2010,Tian2014} therein). For example, Ba$_2$La$_2$NiTe$_2$O$_{12}$ shows an AFM ground state below $T_\mathrm{N} \approx$ 10\,K~\cite{saito2019}, while Ba$_2$La$_2$NiW$_2$O$_{12}$
is ferromagnetically ordered below $T_\mathrm{C} \approx$ 6.3\,K~\cite{yu2023}. Moreover, field-induced up-up-down (UUD) states and magnetization plateaus have been observed in the spin-1/2 or spin-1 triangular-lattice antiferromagnets 
Ba$_2$La$_2$$B$Te$_2$O$_{12}$ ($B$ = Co, Ni)~\cite{kojima2018,saito2019}.

Since the Ba$_2$La$_2$$B$$B'_2$O$_{12}$ family of materials exhibits rich quantum magnetic properties~\cite{kojima2018,yu2023,saito2019,rawl2017,doi2017,park2024},
it is of interest to investigate its last member, namely, 
Ba$_2$La$_2$MnTe$_2$O$_{12}$ (BLMTO) which, to the best
of our knowledge, has not yet been synthesized or studied.
In this paper, we report the successful synthesis of BLMTO and the systematic study of its magnetic properties by 
combining nuclear magnetic resonance (NMR), neutron scattering, and muon-spin spectroscopy ({\textmu}SR) techniques. 
The ground state of the triangular-lattice $XXZ$ model (TL-$XXZ$) was constructed with the tensor network approach to capture the magnetic properties of BLMTO.
%based on the projected entangled simplex state (PESS) ansatz in the thermodynamic limit~\cite{Xie2014}.
Our key observation, the presence of strongly frustrated two-dimensional (2D) magnetism in three-dimensional (3D) materials with a magnetic triangular lattice, suggests that BLMTO and related compounds represent a remarkable platform to study the dimensionality effects on the exotic magnetic states. The hallmark of ideal 2D magnetism within the $ab$-plane of BLMTO — the suppression of long-range order — is manifested here through the magnetic disorder along the $c$-axis.

	\section{Experimental details\label{sec:details}}\enlargethispage{8pt} 

\tcr{A polycrystalline sample of BLMTO was first synthesized by solid-state reaction methods. Stoichiometric amounts of BaCO$_3$ (99.99 \%), La$_2$O$_3$ (99.99 \%), MnCO$_3$ (99.99 \%), and TeO$_2$ (99.999 \%) were mixed and ground thoroughly for 2\,h. The mixture was placed into an alumina crucible and sintered at 1050$^\circ$C for 24\,h. After grinding the samples again, the powders were pressed into pellets and sintered at 1050$^\circ$C for extra 24\,h. Heat-capacity and magnetization measurements were performed on a Quantum Design physical property measurement system and magnetic property measurement system, respectively. The high-field magnetization was measured using an induction method with a multilayer pulse magnet at the Wuhan National Pulsed High Magnetic Field Center and a home-built vibrating sample magnetometer at the Anhui Key Laboratory of Low-Energy Quantum Materials and Devices. To capture the magnetic properties of the BLMTO powder, the ground state of the TL-$XXZ$ model was constructed using the tensor network approach based on the projected entangled simplex state (PESS) ansatz in the thermodynamic limit~\cite{Xie2014}. The NMR measurements were performed on $^{139}$La nuclei ($\gamma_\mathrm{n}= 6.014$\,MHz/T, $I = 7/2$). The frequency-swept spectrum was obtained by integrating the spectral weight at different frequencies. The spin-lattice relaxation rate was measured by the inversion-recovery method, and evaluated by fitting the time-dependent nuclear magnetization to the standard relaxation function for $I = 7/2$. High-resolution neutron powder diffraction (NPD) measurements were carried out on the HRPT powder diffractometer at the Swiss Spallation Neutron Source (SINQ) of the Paul Scherrer Institut (PSI) in Switzerland~\cite{Fischer2000}. The {\textmu}SR measurements were carried out at the general-purpose surface-muon (GPS) instrument at the $\pi$M3 beam line of the Swiss muon source (S{\textmu}S) at PSI in Villigen, Switzerland.  For the {\textmu}SR measurements, a $\sim 1$-mm thick pellet with a diameter of $\sim$8\,mm was positioned on a copper plate using diluted GE varnish, which ensures thermalization at low temperatures.	
All the {\textmu}SR spectra were collected upon heating the sample and were analyzed by means of the musrfit software package~\cite{Suter2012}. All the measurements reported in this paper were performed on the same batch of BLMTO polycrystalline samples.}

\section{\label{sec:results}Results and discussion}\enlargethispage{8pt} 
%=== begin figure ==========================%
\begin{figure}[!htb]
	\centering
	\includegraphics[width=0.48\textwidth,angle=0]{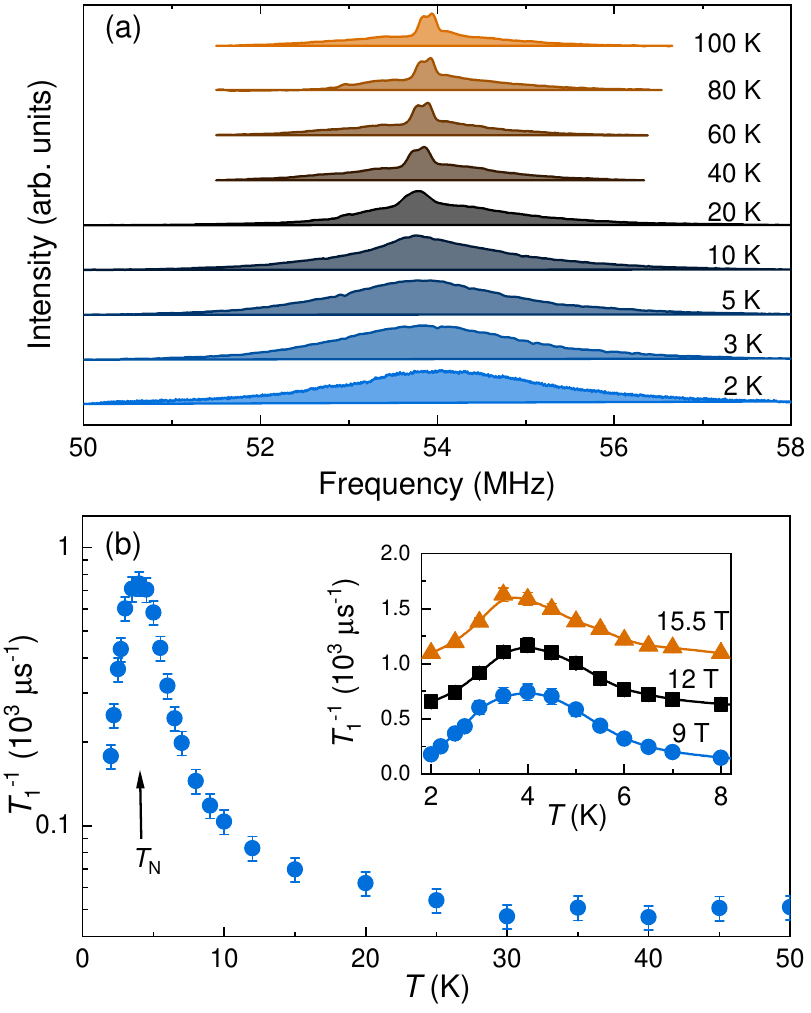}
	\vspace{-2ex}%
	\caption{\label{fig:NMR}%
	(a) $^{139}$La-NMR spectra of BLMTO collected at different temperatures in an applied magnetic field of 9\,T. The NMR spectra in other fields show similar features.
	(b) Spin-lattice relaxation rates $T_1^{-1}$ of BLMTO vs temperature, measured at 9\,T. The inset highlights the mostly unchanged $T_1^{-1}$ peak with the magnetic field. For clarity, the  $T_1^{-1}$ values are shifted vertically.
	}
\end{figure}
%=== end figure ==========================%

The crystal structure of BLMTO was checked by NPD measurements. Refinements of the NPD patterns confirmed the hexagonal structure of BLMTO [Fig.~\ref{fig:Cp}(a)] with a space group $R\bar{3}$ (No.~148) (see Fig.~S1 and Table~S2 in the Supplementary Materials)~\cite{Supple}. 
A tiny amout of extra La$_2$TeO$_6$ (3.45\%) and Ba$_{1/3}$La$_{2/3}$MnO$_3$ (0.36\%) spurious phases were identified, the former was also deteced by x-ray diffraction (see Fig.~S2~\cite{Supple}). \tcr{In addition, the cations at the $A$ and $B$/$B'$ sites are well ordered in BLMTO, with no detectable antisite disorder.}
\tcr{As shown in Fig.~\ref{fig:Cp}(d), the temperature-dependent magnetic susceptibility $\chi(T)$ exhibits a distinct peak at $T_\mathrm{N}$ = 3\,K, corresponding to an AFM transition in BLMTO. The inset shows the inverse susceptibility versus temperature, where the solid red line is a fit to the Curie-Weiss law, which yields an effective moment {\textmu}$_\mathrm{eff}$ = 6.2(2)~{\textmu}$_\mathrm{B}$ and a Curie-Weiss temperature $\theta_\mathrm{CW}$ = -62.4(5)\,K. 
The negative $\theta_\mathrm{CW}$ indicates the presence of dominant AFM interactions in BLMTO, while the effective moment suggests a high spin state of Mn$^{2+}$ ions ($S = 5/2$, {\textmu}$_\mathrm{eff}$ = 5.9~{\textmu}$_\mathrm{B}$). Similar results were obtained for the  $\chi(T)$ data collected under various magnetic field up to 7\,T (see Fig.~S3~\cite{Supple}).
}
%\tcr{Fig. \ref{fig:Cp}(d) shows the low-temperature magnetic susceptibility $\chi(T)$ measured under an applied field of 0.05 T. A peak around 3 K indicates the antiferromagnetic transition. The inset displays the inverse susceptibility as a function of temperature. Fitting the high-temperature linear part to the Curie-Weiss law yields a Curie-Weiss temperature $\theta_\mathrm{CW} \sim -61.2(6)$ K and an effective moment $\mu_\mathrm{eff} \sim 6.2(2)$~{\textmu}$_\mathrm{B}$. The negative $\theta_\mathrm{CW}$ indicates dominant antiferromagnetic interactions, and $\mu_\mathrm{eff}$ is close to the theoretical value of 5.9~{\textmu}$_\mathrm{B}$ for Mn$^{2+}$.}
As shown in Fig.~\ref{fig:Cp}(e), the zero-field specific heat $C(T)/T$ of BLMTO exhibits a broad hump at $T \sim$ 6\,K. Such a hump-like anomaly is robust against external magnetic fields (Fig.~S4~\cite{Supple}) and is clearly distinct from the typical $\lambda$-type transition in magnetic materials with a long-range order~\cite{yu2023,kojima2018,rawl2017}. 
%\tcr{In a conventional ordered magnet, the specific heat anomaly is sharp because the entropy associated with the magnetic transition is released over a narrow temperature window near $T_\mathrm{N}$. By contrast, in a highly frustrated magnet, strong geometric frustration prevents the spins from ordering at a well-defined temperature; instead, short-range correlations and spin fluctuations develop over a wide temperature range, leading to a broad hump in $C(T)/T$ instead of a sharp $\lambda$-peak.}
The magnetic entropy $S_\mathrm{m}$ extends to 50\,K, which is almost 10 times higher than the magnetic ordering temperature of BLMTO. $S_\mathrm{m}$ reaches 13.1\,Jmol$^{-1}$K$^{-1}$ at 50\,K, accounting for 88$\%$ of the theoretical value for a $S = 5/2$ system.
These features differ clearly from those of the isostructural (Sr,Ba)$_2$La$_2$NiW$_2$O$_{12}$ and Ba$_2$La$_2$CoTe$_2$O$_{12}$~\cite{rawl2017,yu2023,kojima2018} and reflect the strong magnetic frustrations in BLMTO (see details below). 
%\tcr{Importantly, the broad specific heat hump does not indicate the absence of long-range order; rather, it is a hallmark of a frustrated magnet where the long-range order is preceded by a wide regime of correlated fluctuations. The actual transition to static long-range order is better captured by local probes such as $\mu$SR and NMR, as discussed below.}

The magnetic transition temperature was also determined using weak transverse-field (wTF-) {\textmu}SR measurements. \tcr{In this case, a field of 3\,mT was applied perpendicular to the initial muon-spin direction. In the paramagnetic (PM) state, this field leads to oscillations (see Fig.~S5~\cite{Supple}). In the long-range ordered AFM state, the same 3-mT field is much smaller than the internal fields (see below). Consequently, upon entering the AFM state, the muon spins precess at frequencies reflecting the internal fields at the muon-stopping sites rather than the weak applied field. Consequently, the PM or nonmagnetic (NM) sample fraction  can be determined from the oscillation amplitude of wTF-{\textmu}SR.}
%\tcr{In the paramagnetic state, all implanted muons experience the same magnetic field $B_\mathrm{appl.}$ = 3 mT, thus precess at the same frequency $\gamma_{\mu}B_\mathrm{appl.}$, giving a large oscillation amplitude in the wTF-$\mu$SR asymmetry spectrum. As the temperature approaches $T_\mathrm{N}$, static magnetic order starts to develop in part of the sample. Muons stopping in the ordered regions sense a distribution of internal fields $B_\mathrm{int}$, causing their precession to effectively disappear from the oscillatory signal. Only muons that stop in the remaining paramagnetic (or non-magnetic) fraction continue to precess at the external field frequency. Consequently, the oscillation amplitude of the wTF-$\mu$SR spectra (denoted $A_\mathrm{NM}$) progressively decreases with the onset of long-range order, as shown in Fig.~\ref{fig:Cp}(f).} 
The asymmetry values $A_\mathrm{NM}$ of wTF-{\textmu}SR spectra shown in
Fig.~\ref{fig:Cp}(f) start to decrease near the onset of AFM order.
The temperature evolution of the magnetic volume fraction can be derived from $V_\mathrm{mag}(T) = 1 - A_\mathrm{NM}(T)/A_\mathrm{NM}(T > T_\mathrm{N})$. 
The $V_\mathrm{mag}(T)$ values for BLMTO are summarized in the inset of Fig.~\ref{fig:Cp}(f), where we show the AFM ordering temperature $T_\mathrm{N} = 4.4(2)$\,K, the transition width $\Delta T = 0.6(1)$\,K, and the zero-temperature magnetic volume fraction $V_\mathrm{mag}(0) = 85(1)\%$. 
The slightly reduced magnetic volume fraction is due to either the
presence of \tcb{spurious} phases (see Figs. S1 and S2~\cite{Supple}), or \tcb{to} muons stopping
on the copper sample holder. The latter usually results in a
temperature-independent {\textmu}SR asymmetry. The sharp AFM transition
indicates that the broad hump in the specific heat is indeed related to magnetic frustration, rather than to the polycrystalline nature of BLMTO. 
%In addition, the large $V_\mathrm{mag}(0)$ confirms that BLMTO can be 
%considered as fully magnetically ordered at low temperatures, indicative of a high sample quality.

Figure~\ref{fig:NMR}(a) shows the $^{139}$La-NMR spectra of BLMTO collected at various temperatures in an applied magnetic field of 9\,T. 
%At high temperatures (i.e., $T \ge$ 40\,K), a split central peak with an intensity-modulated background (contributed by multiple satellites) is observed, consistent with the powder line shape of quadrupole nuclei. For $T <$ $T_\mathrm{N}$, the NMR spectrum develops into a broad hump, which reflects a distribution of the internal hyperfine field at $^{139}$La nuclei. 
In the high temperature range (i.e., $T \ge$ 40\,K), the $^{139}$La-NMR
spectra consist of a split central peak, attributed to the transition
between the $+1/2$ and $-1/2$ nuclear spin states, and
a broad featureless background. \tcr{The simulation of the 100-K NMR spectrum (see Fig.~S6~\cite{Supple}) provides a quadrupole resonance frequency $\nu_\mathrm{Q} \approx 1.6$\,MHz. The line splitting of the central peak is attributed to the second-order correction to the central frequency of the quadrupole interaction. The  background signal arises mainly from the satellite lines of crystals whose $c$-axes are perpendicular to the external field. The absence of singularities in the background signal suggests a broad distribution of local electric-field gradients, which could be caused by disorder effects or local lattice distortions.}

%%%%%%%%%%%%%%%%%%%%%%%%%%%%%%%%%%%%%%%%%%%%%%%%%%%%
\begin{figure}[!htp]
	\centering
	\includegraphics[width=0.45\textwidth,angle=0]{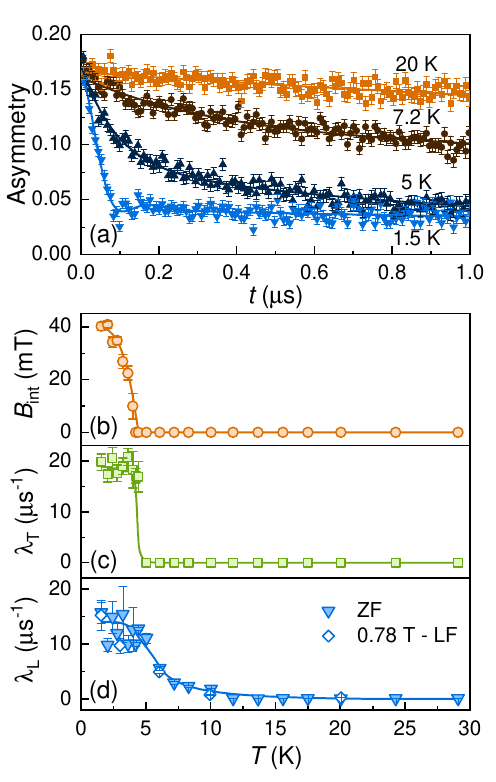}
	%\vspace{-2ex}%
	\caption{\label{fig:musr} \tcr{
		(a) ZF-{\textmu}SR spectra of BLMTO collected at representative
		temperatures. The full-time scale spectra are shown in Fig.~S8, while the decomposed components are presented in Fig.~S9 ~\cite{Supple}.
		Solid lines %through the data
		are fits to Eq.~\eqref{eq:ZF} and Eq~\eqref{eq:ZF22} for the AFM and PM states, respectively. Temperature-dependent internal field $B_\mathrm{int}(T)$ (b), transverse muon-spin relaxation rate $\lambda_\mathrm{T}(T)$ (c), and longitudinal muon-spin relaxation rate $\lambda_\mathrm{L}(T)$ (d). The $\lambda_\mathrm{L}$ values %obtained
		from {\textmu}SR spectra in a longitudinal field of 0.78\,T are also shown in panel (d).  Note that $\lambda_\mathrm{L}$
		is two orders of magnitude larger than
	     $\lambda_\mathrm{tail}$(see Fig.~S9~\cite{Supple}, where also the asymmetries are summarized).
		The solid lines in panels (b)--(d) are guides to the eyes.}
	}
\end{figure}
%%%%%%%%%%%%%%%%%%%%%%%%%%%%%%%%%%%%%%%%%%%%%%%%%%%%%%%%%%

Upon lowering the temperature below 40\,K,
the split peak develops into a single broad peak which, below 20 K, is
barely visible, as it merges with the very broad background contributed
by the satellites. The significant broadening we observe
most likely reflects the formation of the short-range order in BLMTO.  
%Below 40 K, the central peak broadens, loses its split shape by 20 K, and merges with the background by 10 K. This broadening is attributed to the onset of short-range magnetic order and ultimately precluded a clear spectral signature of long-range order. The expected satellite features are not resolved, which could be due to several factors: possible preferential orientation in the polycrystalline sample, strong magnetic anisotropy making the Knight shift highly field-direction sensitive, or inherent structural disorder. As a result, we turned to prove the magnetic phase transition by the temperature dependence of $T_1^{-1}$.} 
%
At high temperatures, the spin-lattice relaxation rate $T_1^{-1}$ is almost temperature independent (see Fig.~S7~\cite{Supple}). It starts to increase below $\sim 25$\,K, to drop again at $T < T_\mathrm{N}$ [see Fig.~\ref{fig:NMR}(b)]. $T_1^{-1}$ measures the weighted sum of the dynamical spin susceptibility at the NMR frequency over the first Brillouin zone, which provides clear signatures of evolving spin correlations. Hence, an increase in $T_1^{-1}$ reflects the enhanced low-energy spin fluctuations as one approaches the magnetic order. Although $T_\mathrm{N}$ shifts slightly with field (to 3.5\,K at 15.5\,T -- see inset), the temperature dependence of $T_1^{-1}(T)$ is quite robust against the external magnetic field. 
%\tcr{The absence of a strong spin-lattice relaxation $T_1^{-1}(T)$ below $T_\mathrm{N}$ in the present NMR data is primarily due to the high applied field (9 T), which suppresses low-energy spin fluctuations, and to the intrinsically different frequency sensitivity of NMR compared to $\mu$SR measurements (to be discussed later).} 

To further investigate the local magnetic properties of BLMTO, a series of zero-field (ZF-) {\textmu}SR spectra were collected at different temperatures, covering both the PM and AFM states. As shown in Fig.~\ref{fig:musr}(a), ZF-{\textmu}SR
spectra in the AFM state (e.g., at 1.5\,K) are characterized by highly damped oscillations, typical of long-range magnetic order, superimposed on a slowly decaying relaxation, observable only at long times. While, in the PM state, ZF-{\textmu}SR spectra still exhibit a relatively fast muon-spin depolarization (e.g., $\sim$2 {\textmu}s$^{-1}$ at 10\,K), implying the presence of strong spin fluctuations. In the AFM state, 
ZF-{\textmu}SR spectra of BLMTO were modeled using the equation: \tcr{
%%%%%%%%%%%%%%%%%%%%%%%%%%%%%%%%%%
\begin{equation}\label{eq:ZF}
	\begin{aligned}
		A_\mathrm{ZF}(t)  &=  A_1  \left[\frac{2}{3} \cos(\omega t + \phi) e^{-\lambda_\mathrm{T} t} + \frac{1}{3} e^{-\lambda_\mathrm{L} t}\right] + A_2 e^{-\lambda_\mathrm{tail} t}\\ 
	&+ A_\mathrm{bg}.
	\end{aligned}
\end{equation}
%%%%%%%%%%%%%%%%%%%%%
}
Here, $A_1$ and $A_2$ represent the asymmetries of the two nonequivalent muon-stopping sites; \tcr{$A_\mathrm{bg}$ represents the background asymmetry, accounting for the muons stopped in the copper sample holder or in the spurious phases. $A_\mathrm{bg}$ was determined to be $\sim$0.02 and was fixed when analyzing the ZF-{\textmu}SR spectra. This background signal corresponds to approximately 11\% of the total asymmetry $A_\mathrm{ZF}$, and is highly consistent with the nonmagnetic volume fraction ($\sim$15\%) determined from the wTF-{\textmu}SR [see Fig.~\ref{fig:Cp}(f)].} 
$\omega$ (= $\gamma_{\mu}$$B_\mathrm{int}$) is the muon-spin precession frequency,  
with $\gamma_{\mu} = 2\pi \times 135.5$\,MHz/T the muon gyromagnetic ratio  
and $B_\mathrm{int}$ the local field sensed by muon spins; $\phi$ is the initial phase;
$\lambda_\mathrm{T}$ and $\lambda_\mathrm{L}$ are the 
transverse and longitudinal muon-spin relaxation rates. In BLMTO, muons stopping at the second site (here, second addend) do not undergo any precession, but show only a slow relaxation, here described by $\lambda_\mathrm{tail}$. 
%\tcr{The constant background asymmetry, $A_\mathrm{bg}$ = 0.020(2), accounts for muons stopping in the copper sample holder or in trace amounts of non‑magnetic impurities. This value was determined from the fit to the lowest‑temperature ZF-$\mu$SR spectrum and then fixed at all temperatures. $\lambda_\mathrm{tail}$ corresponds to approximately 11$\%$ of the total asymmetry and is consistent with the $\sim$15$\%$ reduced magnetic volume fraction obtained from the wTF-$\mu$SR analysis.}
\tcr{In the PM state, the ZF-{\textmu}SR spectra were modeled using the equation:
%%%%%%%%%%%%%%%%%%%%%%%%%%%%%%%%%%
\begin{equation}\label{eq:ZF22}
		A_\mathrm{ZF}(t)  =  A_1 e^{-\lambda_\mathrm{L} t} + A_2 e^{-\lambda_\mathrm{tail} t} + A_\mathrm{bg},
\end{equation}
%%%%%%%%%%%%%%%%%%%%%%%%%%%%%%%%%%
with all the parameters being identical to those in Eq.~\eqref{eq:ZF}.}
%\tcr{Here, $A_1$ and $A_2$ represent the asymmetries of the fast‑relaxing and slow‑relaxing channels, respectively, which is distinct from the fit in AFM state where they denote the asymmetries of two distinct muon stopping sites. The parameters $\lambda_\mathrm{L}$ and $\lambda_\mathrm{tail}$ correspond to the fast and slow relaxation rates, respectively.}

\begin{figure}[!htp]
	\centering
	\includegraphics[width=0.48\textwidth,angle=0]{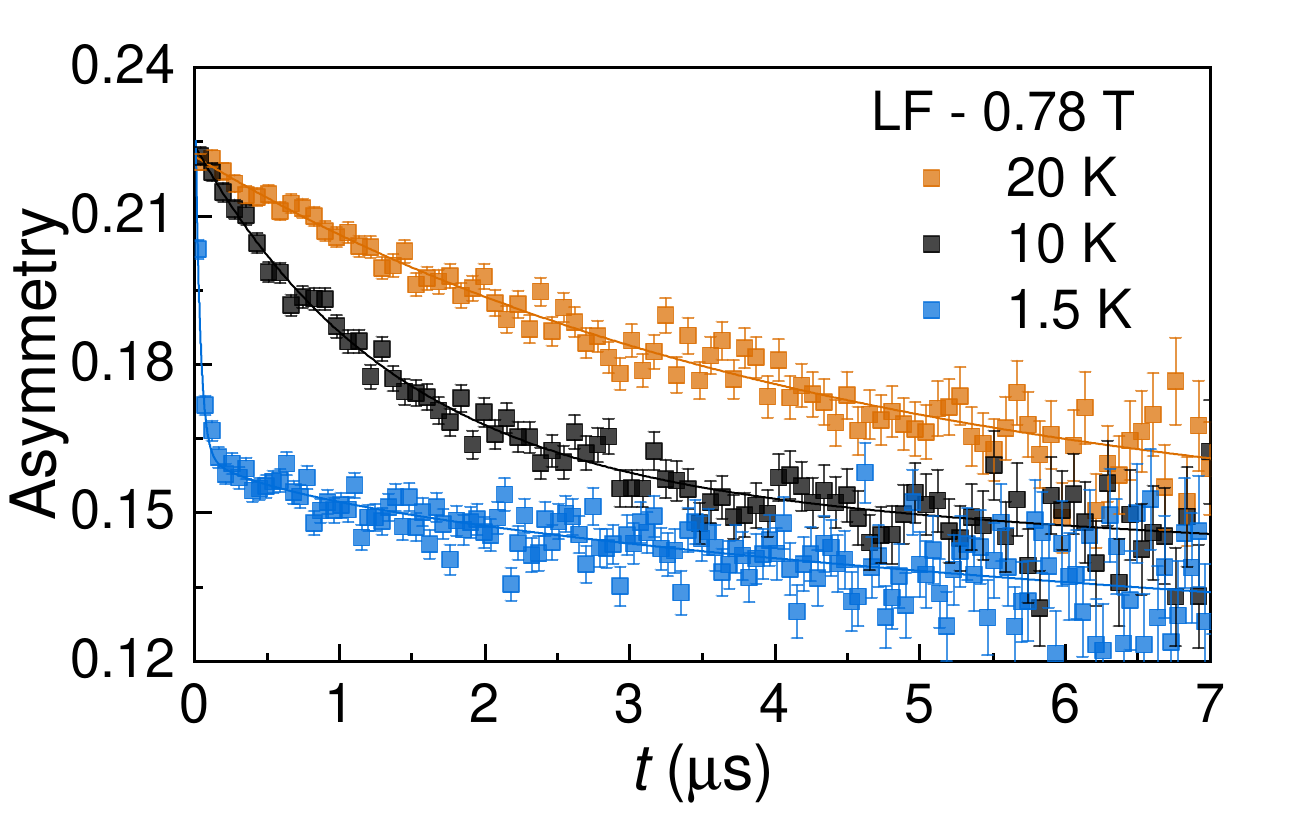}
	%\vspace{-2ex}%
	\caption{\label{fig:LF-musr} \tcr{
		 LF-{\textmu}SR spectra collected in a magnetic field of 0.78\,T at selected 
		temperatures, covering both the AFM and PM states of BLMTO. 
		Solid lines are fits to Eq.~\eqref{eq:ZF22}. The applied magnetic field is parallel to the muon-spin direction. The LF-{\textmu}SR spectra collected at $T = 60$\,K in different magnetic fields are shown in Fig.~S10~\cite{Supple}. 
	}}
\end{figure}

\begin{figure*}[!htp]
	\centering
	\includegraphics[width=1\textwidth,angle=0]{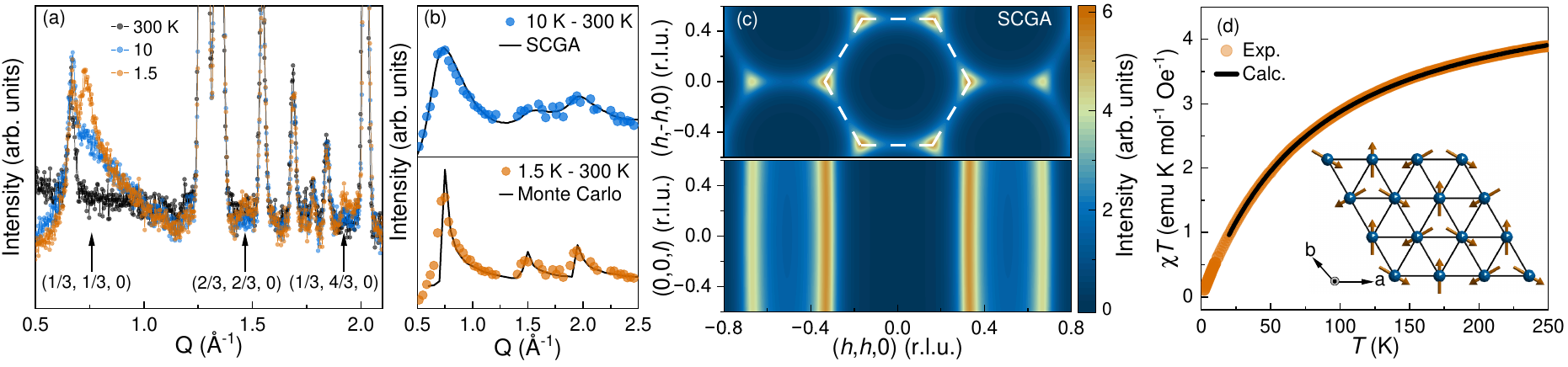}
	\caption{\label{fig:NPDD}(a) NPD patterns collected at $T$ = 1.5, 10, and 300\,K reveal asymmetric magnetic Bragg peaks with $\bm{q}$ =  (1/3, 1/3, 0), (2/3, 2/3, 0), and (1/3, 4/3, 0) (marked by arrows). (b) Comparison of experimental- and simulated NPD spectra. Upper panel: experimental data at 10\,K is fitted using the $J_1$-$J_2$-$J_\mathrm{c}$ model through the SCGA method. Lower panel: experimental data at 1.5\,K and the simulated pattern calculated using the fitted $J_1$-$J_2$-$J_\mathrm{c}$ model through classical Monte Carlo simulations. (c) Simulated diffuse neutron scattering patterns in the ($h$, $k$, 0) and ($h$, $h$, $l$) planes using the fitted $J_1$-$J_2$-$J_\mathrm{c}$ calculated %by the
		from the SCGA method at 10\,K. (d) Temperature dependence of the reduced magnetic susceptibility $\chi$$T$ of BLMTO, measured at 7\,T. The solid line represents the theoretical calculation using the SCGA method.  
		The inset depicts the 120$^{\circ}$ AFM correlations among the Mn$^{2+}$ electronic moments in the $ab$-plane.}
\end{figure*}

Internal local field $B_\mathrm{int}$ increases monotonically as the temperature decreases below $T_\mathrm{N}$, reaching $\sim$ 40\,mT at 1.5\,K [Fig.~\ref{fig:musr}(b)]. Such field is almost 10 times smaller than that observed in the isostructural Ba$_2$La$_2$NiW$_2$O$_{12}$, the latter showing an FM ground state below 6\,K~\cite{yu2023}. $\lambda_\mathrm{T}$ is a measure of the width of the static magnetic field distribution at the muon-stopping site, but it is also affected by dynamic effects, as e.g., spin fluctuations. Conversely, $\lambda_\mathrm{L}$ is determined solely by spin fluctuations. In BLMTO, $\lambda_\mathrm{T}$ is zero in the PM state, but it becomes increasingly prominent and temperature independent at $T < T_\mathrm{N}$ [Fig.~\ref{fig:musr}(c)]. 
Such a large $\lambda_\mathrm{T}$ at temperatures far below $T_\mathrm{N}$ is unusual for an antiferromagnet, and implies a disordered field distribution in the AFM state of BLMTO.
At $T$ = 1.5\,K, $\lambda_\mathrm{T}$ is about $\sim$20\,{\textmu}s$^{-1}$ in BLMTO, which is almost 4 times larger than %that of
in Ba$_2$La$_2$NiW$_2$O$_{12}$~\cite{yu2023}. Such a significantly inhomogeneous field distribution in the AFM state of BLMTO is most likely attributed to its disordered magnetic structure along the $c$-axis (see details below).  
For typical materials with a long-range (anti)ferromagnetic order~\cite{yu2023,Zhu2022,wang2024}, $\lambda_\mathrm{L}$ diverges near $T_\mathrm{N}$, followed by a significant drop at $T < T_\mathrm{N}$, indicating that spin fluctuations are the strongest close to the onset of the magnetic order. 
In contrast, in BLMTO, $\lambda_\mathrm{L}$ increases continuously as the temperature decreases below 12 K [Fig.~\ref{fig:musr}(d)], where also the specific heat shows an upturn [Fig.~\ref{fig:Cp}(e)] thus indicating enhanced spin fluctuations at low temperatures.
At base temperature, $\lambda_\mathrm{L}$ reaches $\sim 16$\,{\textmu}s$^{-1}$, i.e., a comparable value to $\lambda_\mathrm{T}$ [Fig.~\ref{fig:musr}(c)]. Such strong spin fluctuations in the AFM state are quite robust against the external field. \tcr{Figure~\ref{fig:LF-musr} shows the longitudinal-field (LF-) {\textmu}SR spectra collected in a magnetic field of 0.78\,T at temperatures covering both the AFM and PM states of BLMTO. At $T$ = 1.5\,K, the internal field of 40\,mT is significantly smaller than the external longitudinal field of 0.78\,T. As a consequence, muon spins do not precess around the local field. In this case, the LF-{\textmu}SR spectra can also be modeled using Eq.~\eqref{eq:ZF22}. The derived muon-spin relaxation rates $\lambda_\mathrm{L}$ and  $\lambda_\mathrm{tail}$ are summarized in Fig.~\ref{fig:musr}(d) and Fig.~S11~\cite{Supple}, respectively. The $\lambda_\mathrm{L}$ values in a 0.78-T field almost overlap with the zero-field values.
Such robust spin fluctuations are attributed to the disordered magnetic structure along the $c$-axis and the strong magnetic frustration of the Mn$^{2+}$ triangular lattice. By contrast, the weak relaxation (i.e., $\lambda_\mathrm{tail}$) is suppressed by the longitudinal field (see Fig.~S11~\cite{Supple}). Note that, the decrease of $T_1$ at $T < T_\mathrm{N}$ in Fig.~\ref{fig:NMR}(b) indicates a slowing down of spin fluctuations in the AFM state, which appears to contradict $\lambda_\mathrm{L}$ in Fig.~\ref{fig:musr}(d). However, NMR and {\textmu}SR measurements were performed under very different magnetic-field conditions: the former in fields of 9, 12, and 15.5\,T and the latter in fields of 0 and 0.78\,T. Although muon-spin relaxation is robust against longitudinal fields up to 0.78\,T, it can be suppressed at larger magnetic fields, where the spin fluctuations may also be suppressed. Further {\textmu}SR investigations
are required to clarify this behavior as a function of the applied field.}

%\tcr{The strong spin fluctuations exhibit remarkable robustness against an external magnetic field, with the relaxation rate $\lambda_\mathrm{L}$ remaining nearly unchanged even under a field of 0.78 T compared to the zero-field condition. The corresponding LF-$\mu$SR spectra and analysis are presented in Supplementary Materials S7-S8 and Note S4~\cite{Supple}. This robust fluctuations are naturally attributed to the disordered magnetic structure along the $c$-axis and the strong magnetic frustration of the Mn$^{2+}$ triangular lattice.}

%	Such strong spin fluctuations in the AFM state are quite robust against the external field \tcr{[see longitudinal-field (LF-) {\textmu}SR results in Figs.~\ref{fig:musr}(d), Figs.~S7-S8, and Note S4~\cite{Supple}]} and are attributed to the disordered magnetic structure along the $c$-axis and the strong magnetic frustration of the Mn$^{2+}$ triangular lattice.

Neutron scattering measurements were employed to reveal the magnetic structure of BLMTO. As shown in Fig.~\ref{fig:NPDD}(a), NPD patterns exhibit a distinct broad hump near $\bm{q} = (1/3,1/3,0)$ as the temperature approaches 10\,K.  
Such a hump reflects the emergence of short-range order already above $T_\mathrm{N}$, leading to a highly asymmetric magnetic peak at $Q \sim 0.75$~\AA$^{-1}$ at 1.5\,K. The asymmetric magnetic peak suggests the presence of two-dimensional magnetic correlations in BLMTO~\cite{PhysRevB.89.224511}.
To quantitatively analyze the spin correlations, we employ complementary theoretical approaches in the different temperature regimes. Above $T_\mathrm{N}$ (10\,K), the self-consistent Gaussian approximation (SCGA) method~\cite{conlon2010} successfully  models the diffuse scattering [Fig.~\ref{fig:NPDD}(b), upper panel] and the temperature dependence of the reduced magnetic susceptibility, $\chi T$ [Fig.~\ref{fig:NPDD}(d)], using an antiferromagnetic triangular-lattice model (AFTL) with a fitted coupling strength of 
$J_1 = 0.323(6)$, $J_2 = 0.028(1)$, and $J_{\mathrm{c}} = -0.007(4)$ meV. Here, $J_1$, $J_2$, and $J_{\mathrm{c}}$ represent the nearest-neighbor, next-nearest-neighbor coupling in the $ab$-plane, and the interlayer coupling, respectively. Notably,
%The fitted value of
an almost zero $J_{\mathrm{c}}$ value confirms the 2D character of BLMTO. Below $T_\mathrm{N}$ (1.5 K), classical Monte Carlo simulations were utilized to capture the quasi-long-range order within the triangular layers. As shown in the lower panel of Fig.~\ref{fig:NPDD}(b), the asymmetric peaks centered around $\bm{q}$ = (1/3, 1/3, 0), (2/3, 2/3, 0), and (1/3, 4/3, 0) are well reproduced by the fitted AFTL model. As shown in  Fig.~\ref{fig:NPDD}(c), diffuse scattering calculated using the SCGA method at 10 K suggest that the scattering intensity is centered around the $K$-(1/3, 1/3, $l$) points regardless of the $l$ value. This key fact reveals two-dimensional spin fluctuations that form a typical 120$^{\circ}$ order in the $ab$-plane, but remain disordered along the $c$-axis, as has been previously observed in other triangular-lattice compounds~\cite{Ding2018b,Xing2019,Kojima2022,Fritsch2017}.
%is quite common in the quasi-2d triangular-lattice compounds 

Previous neutron studies of the isostructural Ba$_2$\-La$_2$\-$B$\-Te$_2$\-O$_{12}$ ($B$ = Co, Ni) compounds have proposed magnetic structures based solely on background-subtracted diffraction patterns, without a detailed analysis~\cite{kojima2018,saito2019,park2024}. 
In particular, the nearly degenerate (1/3, 1/3, 0) and (1/3, 1/3, 1) magnetic peaks in these materials can hardly distinguish between 3D ordered states and 2D magnetism.  
By combining the SCGA method with Monte-Carlo simulations, we establish an exotic magnetic ground state that is better in accord with the materials' dimensional constraints, namely, 120$^{\circ}$ AFM order within the $ab$-plane, but magnetic disorder along the $c$-axis. 
\tcr{Such a unique magnetic order reflects a quasi-static magnetic ground state of BLMTO, which is consistent with the persistent muon-spin relaxation rate in its AFM state [see Fig.~\ref{fig:musr}(d)].}
The revised 2D-like magnetic structure not only challenges the interpretation of 3D-ordered states in $A_4$$B$$B_2'$O$_{12}$ family of materials, but it also provides further insight into how the dimensionality affects the exotic magnetic states.

\section{\label{ssec:diss} Discussion}   

The triangular lattice usually promotes a large magnetic frustration~\cite{paddison2017,li2020,cheng2011,zhou2012,gao2022,sheng2025,bordelon2019,xiang2024,anderson1973,collins1997,shirata2012,starykh2015},
often quantified by the ratio $f$ = |$\theta_\mathrm{CW}$|/$T_\mathrm{N}$, with $\theta_\mathrm{CW}$ the Curie-Weiss temperature, as estimated from the temperature-dependent magnetic susceptibility. For BLMTO, the $f$ = 14.2 exceeds the threshold value ($f$ = 10) for strongly frustrated magnets, and is much larger than that of other frustrated magnets with a long-range order (see \tcb{Table~\ref{tab:family}).}
%While this finite value is indeed smaller than the infinite $f$ of QSL candidates, it is larger than that of most magnetically ordered frustrated materials (see Table Table~\ref{tab:family}). 
Such a significant magnetic frustration leads to persistent spin fluctuations in the magnetically ordered state of BLMTO, reflected in continuously increasing of muon-spin relaxation rates [Fig.~\ref{fig:musr}(d)]. Indeed, $\lambda_\mathrm{L}$ of BLMTO is almost 10 times larger than that of most QSL candidates.
% Thus, significant magnetic frustration places BLMTO in the rare regime between conventional frustrated antiferromagnets and QSLs.} %implying %the presence of %\tcr{persistent} magnetic fluctuations in its ground state. 
In addition to the easy-axis anisotropy, such fluctuations may also lead to a 1/3-saturation-magnetization ($M_\mathrm{s}$/3) plateau in many frustrated magnets, corresponding to an intermediate UUD phase~\cite{alicea2009,zhou2012,saito2019,kojima2018,li2020,shu2023}. 

%%%%%%%%%%%%%%%%%%%%%%%
\begin{figure}[!htp]
	\centering
	\includegraphics[width=0.45\textwidth,angle=0]{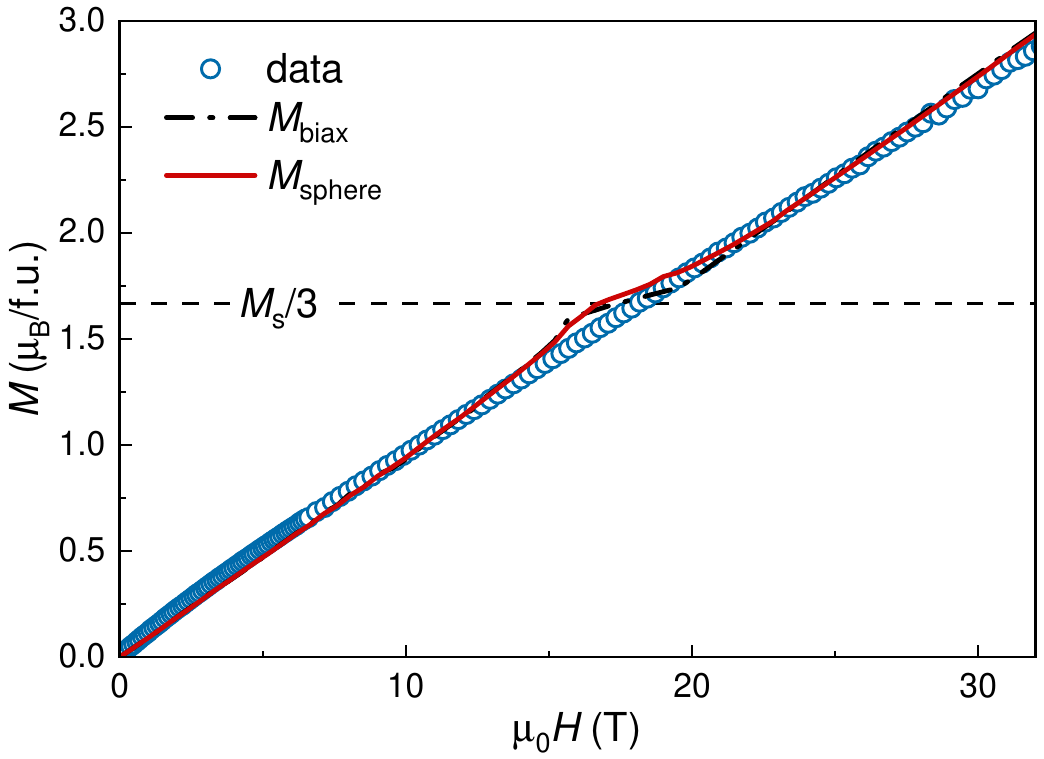}
	\caption{\label{fig:MH} Field-dependent magnetization $M(H)$ collected at $T = 1.8$\,K in magnetic fields up to 32\,T (data up to 49\,T are presented in Fig.~S12~\cite{Supple}). %\tcr{The} derivative with respect to field, d$M$/d$H$, is also shown (right-axis). 
	Red solid and black dash-dotted lines represent the calculated magnetization using spherical ($M_\mathrm{sphere}$) and weighted-biaxial averages ($M_\mathrm{biax}$), respectively. The dashed horizontal line denotes the $M_\mathrm{s}$/3 magnetization plateau. 
	}
\end{figure}
%%%%%%%%%%%%%%%%%%%%%%%

%%%%%%%%%%%%%%%%%%%%%%
\begin{table}[!bht]
	\centering
	\caption{\label{tab:family} Magnetic frustration ratio $f$ and ZF muon-spin relaxation rates $\lambda_\mathrm{L}$ at base temperature for various types of magnets.} 		
	\begin{tabular}{lcccl}
		\toprule
		\textrm{Compound}&
		\textrm{Magnetic lattice}&
		\textrm{$f$}&
		\textrm{$\lambda_\mathrm{L}$ ({\textmu}s$^{-1}$)}&
		\textrm{Ref.}\\
		\colrule
		ZnCu$_3$(OH)$_6$Cl$_2$        & Kagome    & $\infty$    & 0.5 &  \cite{helton2007,fak2012} \\
		Tm$_3$Sb$_3$Zn$_2$O$_{14}$    & Kagome    &  $\infty$   & 1.3 &  \cite{ding2018} \\
		MgFe$_3$(OH)$_6$Cl$_2$        & Kagome    &  3.9  & 0.08 &  \cite{fujihala2017} \\
		H$_3$LiIr$_2$O$_6$            & Honeycomb     &  $\infty$   & 4.3 &  \cite{yang2024} \\
		Na$_2$IrO$_3$                  & Honeycomb     &  8.3  & 10 &   \cite{singh2012} \\
		$\alpha$-RuCl$_3$             & Honeycomb     &  2.9  & 3.5 &  \cite{sears2015,lang2016} \\
		YbMgGaO$_4$                   & Triangular    &  $\infty$   & 0.3 &  \cite{li2016} \\
		NaYbO$_2$                     & Triangular    &  $\infty$   & 0.9 &  \cite{bordelon2019,ding2019} \\  
		Ba$_2$La$_2$NiW$_2$O$_{12}$   & Triangular    &  4.1  & 0.53&  \cite{yu2023}  \\
		Ba$_2$La$_2$MnTe$_2$O$_{12}$  & Triangular    &  14.2 & 16  & This work     \\   \bottomrule		
	\end{tabular}
	\\ %\vspace{8pt}
\end{table}
%%%%%%%%%%%%%%%%%%%%%%%%%%%%

To investigate the effects of magnetic fluctuations, the field-dependent magnetization $M(H)$ of BLMTO was measured at 1.8\,K in magnetic fields up to 49\,T. As shown in Fig.~\ref{fig:MH}, the magnetization is almost linear in field, confirming the absence of magnetization plateaus and UUD phases in BLMTO. Since both $J_2$ and $J_\mathrm{c}$ are much weaker than $J_1$, we employ the nearest-neighbor (NN) TL-$XXZ$ model to calculate the magnetization of BLMTO.  
This model 
%exhibits a variety of quantum phases, including the Y, V, UUD, and umbrella phases, etc~\cite{kojima2018,yamamoto2014,saito2019,park2024}, and
is described by the Hamiltonian:
%%%%%%%%%%%%%%%%%%%%%%
\begin{equation}
	\mathcal{H} = J \sum_{\langle i,j \rangle} (S_i^x S_j^x + S_i^y S_j^y + \eta S_i^z S_j^z) - h_x \sum_i S_i^x - h_z \sum_i S_i^z.
	\label{eq:XXZ}
\end{equation}
%%%%%%%%%%%%%%%%
Here, $J > 0$ is the AFM exchange coupling, $S_i^{x,y,z}$ are the spin operators at the $i$-th site of 
the triangular lattice; $\langle i, j \rangle$ denotes the NN pairs; $\eta$ ($<$ 1) suggests an easy-plane anisotropy; $h_{x}$ and $h_{z}$ represent the reduced magnetic fields within the $ab$-plane or along the $c$-axis, respectively.
%The triangular lattice was decomposed into three sublattices, each assigned a local tensor that constitutes an infinite PESS as a wave function converging to the ground state through imaginary-time evolutions, with bond dimension of four
The ground state wave function of the system is obtained using the PESS method (Fig.~S13 and Note~S1~\cite{Supple}). After convergence, the expectation values of the spin operator are calculated to determine the magnetization $M_x$ and $M_z$ induced by the fields $h_x$ and $h_z$, respectively. Assuming the susceptibility can be decomposed into constant components
along both the $x$- and $z$-directions, the magnetization of powder samples can be evaluated using a biaxial average, namely, $M_\mathrm{biax} = (2M_x + M_z)/3$.
As shown by the black dash-dotted line in Fig.~\ref{fig:MH}, by using the $J$ ($= J_1$) value determined from NPD (see Fig.~\ref{fig:NPDD}) and assuming $\eta = 0.9$, the calculated $M_\mathrm{biax}$ agrees reasonably well with the experimental data. 
However, $M_\mathrm{biax}$ exhibits a slight bump around 
$M_\mathrm{s}$/3, here attributed to the stabilization of
a UUD phase by $h_x$ for a wide range of $\eta$-s (see Fig.~S14 and Note~S2~\cite{Supple}). This indicates that the NN TL-$XXZ$ model serves only as an approximate effective model for BLMTO. The complete vanishing
of the plateau requires taking into account additional, more complex interactions, which would greatly increase the computational complexity of the tensor network approach. Nevertheless, the NN TL-$XXZ$ model is a good base
for further improvements. For polycrystalline (powder) materials, the applied magnetic field acts only as an average effect, leading to a rather coarse biaxial averaging of the magnetization. Therefore, we further performed a spherical averaging (see Fig.~S15 and Note S2~\cite{Supple}), obtaining a weaker magnetization bump around $M_\mathrm{s}$/3 (Fig.~\ref{fig:MH}).

The above discrepancy can be understood by considering the following factors. In general, the width of the $M_\mathrm{s}$/3
magnetization plateau is determined by both the magnetic exchange coupling and the effective spin, namely, $\Delta H = 1.8J/2S$~\cite{chubukov1991,alicea2009}. For Mn$^{2+}$ ($S = 5/2$), its relatively large effective spin renders the $M_\mathrm{s}$/3 plateau less distinct compared to Co$^{2+}$ ($S = 1/2$) and Ni$^{2+}$ ($S = 1$) ions with small effective spin~\cite{kojima2018,saito2019}. Different from the easy-\emph{axis} anisotropy, that generally stabilizes the UUD phase, significant quantum fluctuations are needed to maintain such
phase in easy-\emph{plane} anisotropy compounds~\cite{chubukov1991,yamamoto2014,sellmann2015}, such as BLMTO. In addition, in the $XXZ$ model, other magnetic interactions beyond NN %ones
might need to be considered.
For example, the next-nearest-neighbor (NNN) interaction favors the stripe phase over the UUD phase in a Heisenberg antiferromagnet on a frustrated triangular lattice~\cite{Ye2017}. Such a NNN interaction can also result in a less distinguishable $M_\mathrm{s}$/3 plateau~\cite{chubukov1992,nakano2017}, resembling the results of BLMTO (see Fig.~\ref{fig:MH}). Although the estimated $J_2$ = 0.028\,meV is much smaller than that of $J_1$ = 0.323\,meV, its presence might affect the magnetization plateau in BLMTO.
Given that the NNN interaction significantly increases the degrees of freedom of the minimum unit to describe the whole lattice system, alternative approaches, such as spin-wave theory, may provide further insights.

\section{\label{ssec:Sum}Conclusion}

In summary, a new member of the hexagonal perovskites $A_4BB'_2$O$_{12}$ with magnetic triangular lattice, namely,  Ba$_2$La$_2$MnTe$_2$O$_{12}$, was synthesized, and its magnetic properties were systematically studied by a variety of techniques. 
BLMTO undergoes a magnetic transition at $T_\mathrm{N}$ $\sim$ 4.4\,K, %which is
attributed to the AFM order of the Mn$^{2+}$ sublattice. The NPD measurements reveal an exotic magnetic ground state in BLMTO, where manganese moments from a 120$^{\circ}$ AFM order within the $ab$-plane, but are disordered along the $c$-axis. Such a magnetic state is highly consistent with the persistently strong spin fluctuations and large internal field distributions in the ZF-{\textmu}SR spectra even %at temperatures
below $T_\mathrm{N}$. 
BLMTO represents a rare case of a 3D hexagonal perovskite %but
hosting a strongly frustrated 2D magnetism.
Such a revised 2D magnetism challenges the interpretation of the 3D ordered states in the $A_4$$B$$B_2$'O$_{12}$ family of materials.
Though BLMTO exhibits strong magnetic frustration and fluctuations, their effects are largely suppressed %due to
by the high-spin state of Mn$^{2+}$ and the easy-plane anisotropy, both of which are unfavorable to a quantum phase stabilization.
Future studies, including single-crystal growth and chemical substitutions on $A$ or $B$' sites, are highly desirable to search for possible quantum phase transitions in BLMTO.

\vspace{1pt}
\begin{acknowledgments}
This work was supported by the National Natural Science Foundation of China (Grant Nos.~12374105, 12350710785, and 12561160109), and the Fundamental Research Funds for the Central Universities. 
H.Z.\ and S.L.\ acknowledge the support from the National Natural Science Foundation of China via Grant No.\ 12274126, while J.M.\ was supported via Grant Nos.\ U2032213 and 12334008.
%acknowledges the financial support from the National Natural Science Foundation of China
W.W.\ and S.G.\ acknowledge financial support from the National Science Foundation of China (Grant No.\ 12374152). 
Z. X.\ acknowledge the support the National Key Research and Development Program of China (Grant No.~2023YFA1406500).
Neutron powder diffraction and muon-spin spectroscopy measurements were performed at the Swiss spallation neutron source SINQ and Swiss muon source (S{\textmu}S) of the Paul Scherrer Institut (PSI) in Switzerland. We also acknowledge the allocation of beam time at HRPT diffractometer and GPS spectrometer.

\end{acknowledgments}

\bibliography{BLMTO}
%\begin{footnotesize}
%\end{footnotesize}
%\printbibliography
\end{document}